\documentclass[a4paper,10pt,reqno]{article}
\usepackage{ae} 
\usepackage[T1]{fontenc}
\usepackage[ansinew]{inputenc}
\usepackage{amsmath}
\usepackage{amssymb}
\usepackage{graphicx}
\usepackage[dvipdfm,usenames,x11names,svgnames,dvipsnames]{xcolor}
\usepackage{cancel}

\usepackage{tikz}
\usepackage[normalem]{ulem}
\usetikzlibrary{arrows,shapes}

\newcommand{\vb}[1]{{\boldsymbol {#1}}}

\newcommand{\ud}{\mathop{}\!\mathrm{d}} 

\title  {
            On the Green-Functions of the \\
            classical offshell electrodynamics \\
            under the manifestly covariant relativistic
            dynamics of Stueckelberg
        }
\author{ I. Aharonovich$^{a}$
         and
         L. P. Horwitz$^{abcd}$ \\
         \\
         {\scriptsize $^{a}$ Bar-Ilan University, Department of Physics, Ramat Gan, Israel.     }  \\
         {\scriptsize $^{b}$ Tel-Aviv University, School of Physics, Ramat Aviv, Israel.        }  \\
         {\scriptsize $^{c}$ College of Judea and Samaria, Ariel, Israel.                       }  \\
         {\scriptsize $^{d}$ IYAR, Israel Institute for Advanced Research, Rehovot, Israel      }
       }

\begin{document}
\maketitle
\begin{abstract}
    In previous papers
    derivations 
    of the Green function  
    have been given
    for
    5D off-shell electrodynamics in the framework of the manifestly covariant 
    relativistic dynamics of Stueckelberg (with invariant evolution parameter $\tau$).
    In this paper, we reconcile these derivations resulting in different explicit forms, and
    relate our results to the conventional fundamental solutions of linear 5D wave equations
    published in the mathematical literature.
    We give physical arguments for the choice of the Green function retarded in the fifth variable $\tau$.
\end{abstract}

\maketitle

\section{Introduction}
Classical 5D electrodynamics arises as a $U(1)$ gauge of the relativistic quantum mechanical Schr\"{o}dinger
equation \cite{Horwitz1993,Horwitz1998,LandHor1991,LandShnerbHorwitz1995,SaadHorArsh1989}, similar to the construction
of Maxwell fields from the $U(1)$ gauge of the classical Schr\"{o}dinger equation.

We have studied the configuration of such fields associated with a uniformly moving source 
\cite{horjig2006} as well as from a uniformly accelerating one \cite{aharonovich_2009}.
The action of the resulting generalized Lorentz force on the source 
(radiation reaction) is under study; the results, very different in nature from the Abraham-Lorentz-Dirac analysis
(e.g., \cite{dirac_1938, Rohrlich1990}), will be reported elsewhere \cite{aharonovich_2011}.

By requiring local gauge invariance of 
\begin{align}
    \label{eq:stueckelberg_hamilton_equation}
    i \dfrac{\partial}{\partial \tau} \psi_{\tau}(x) & = \dfrac{1}{2M} p^{\mu} p_{\mu} \psi_{\tau}(x),
\end{align}
where $p^{\mu}$ is represented by $- i \partial/\partial x^{\mu}$,      
five compensation fields are introduced \cite{Horwitz1993,Horwitz1998,LandHor1991,LandShnerbHorwitz1995}.

Under the 5D generalized Lorentz gauge, these fields obey a 5D wave equation of the form
($\eta_{\mu\nu} = -, +, +, +$)
\begin{align}
    \label{eq:5d_wave_equation}
    \left(
        \eta^{\mu \nu}
        \dfrac{\partial}{\partial x^{\mu}}
        \dfrac{\partial}{\partial x^{\nu}}
        + 
        \sigma_{55}
        \dfrac{\partial^2}{\partial \tau^2}
    \right)
    a^{\alpha}(x,\tau)
    & \equiv
        \partial_{\beta} \partial^{\beta} a^{\alpha}(x,\tau)
    =
        j^{\alpha}(x,\tau)
\end{align}
where $x = (x^{\mu}) = (t, x^{i})$ is a 4D spacetime coordinate and $\alpha,\beta \in \{0,1,2,3,5\}$ run over the entire 5D coordinates.
Here, $x^{5} \equiv \tau$, whereas $\mu,\nu \in \{0,1,2,3\}$ run over 
the 4D spacetime coordinates;
$\sigma_{55} = \pm 1$ is the signature of $\tau$ coordinate in the wave equation, denoting either $O(4,1)$ 
or $O(3,2)$ symmetry of the homogeneous wave equation.

We shall use $\sigma_{55} = +1$ (corresponding to $O(4,1)$)  here, although most of the results can easily be extended to the $\sigma_{55} = -1$
case as well.

The Green function (GF) associated with \eqref{eq:5d_wave_equation} obeys the equation
\begin{align}
    \label{eq:5d_green_function_equation}
    \partial_{\beta} \partial^{\beta} g(x,\tau) & = \delta^{4}(x) \delta(\tau)
\end{align}

There are numerous ways to solve \eqref{eq:5d_green_function_equation} without referring directly 
to the Fourier transform; most of these involve using the $O(4,1)$ symmetry of the equation.

Nevertheless, in the works of Land and Horwitz \cite{LandHor1991} and Oron \emph{et al.} \cite{OronHorwitz2000} mentioned above, 
the Fourier method was used, for which $g(x,\tau)$ is represented by
\begin{align}
    \label{eq:5d_green_function_fourier_transform}
    g(x,\tau)
    & = 
        \dfrac{1}{(2\pi)^5}
        \int_{\mathbb{R}^{5}}
            \ud^4k 
            \,
            \ud k_{5}
            \,
            \dfrac{     e^{i (k_{\mu} x^{\mu} + k_{5}\tau) }            }
                  {     k_{\mu} k^{\mu} + k_{5}^2                       }
    =
        \dfrac{1}{(2\pi)^5}
        \int_{\mathbb{R}^{5}}
            \ud^5k 
            \,
            \dfrac{     e^{i (k_{\alpha} x^{\alpha}      ) }            }
                  {     k_{\alpha} k^{\alpha}                           }
\end{align}

Solutions of \eqref{eq:5d_green_function_fourier_transform} were obtained in
\begin{itemize}
 \item Land and Horwitz \cite{LandHor1991}, using an integral over Schwinger's result \cite{PhysRev.82.664}
       obtaining
        \begin{align}
            \label{eq:land_green_function}
            g_{P}(x,\tau) 
            & = 
                - 
                \dfrac{1}{4\pi} \delta(x^2) \delta(\tau)
                -
                \dfrac{1}{2\pi^2}
                \dfrac{\partial}{\partial x^2}
                \begin{cases}
                    \dfrac{\theta(x^2 - \tau^2)}
                          {\sqrt{x^2 - \tau^2} }
                    \qquad 
                    &
                    O(3,2)
                \\
                    \dfrac{\theta(-x^2 - \tau^2)}
                          {\sqrt{-x^2 - \tau^2} }
                    \qquad 
                    &
                    O(4,1)
                \end{cases}
        \end{align}
        where $g_{P}$ refers to the \emph{Principal Part} solution, and $x^2 = x_{\mu} x^{\mu} = r^2 - t^2$.
       
 \item Oron and Horwitz \cite{OronHorwitz2000}, integrating first using $k_{5}$,
       in which the result obtained is (for $O(4,1)$):
        \begin{align}
            \label{eq:ori_green_function}
            g(x,\tau)
            & = 
                \dfrac{2 \theta(\tau)}
                      {(2\pi)^3      }
                \times
                \begin{cases}
                    {
                        \scriptsize
                        \frac{1}{[- x^2 - \tau^2]^{3/2}}
                        \tan^{-1}
                        \left(
                            \frac{1}{\tau}
                            \sqrt{-x^2 - \tau^2}
                        \right)
                        - 
                        \frac{\tau}{x^2(x^2 + \tau^2)}
                    }
                \vspace{3mm}
                \\
                    {
                        \scriptsize
                        \frac{1}{2}
                        \frac{1}{[x^2 + \tau^2]^{3/2}}
                        \ln
                        \Big|
                            \frac{ \tau - \sqrt{\tau^2 + x^2}}
                                 { \tau + \sqrt{\tau^2 + x^2}}
                        \Big|
                        - 
                        \frac{\tau}{x^2(x^2 + \tau^2)}
                    }
                \end{cases}
        \end{align}

 \item Aharonovich and Horwitz \cite{horjig2006} resulting in equation \eqref{eq:green_function_expected_solution} (below). 
       Using a different method, a \emph{$\tau$-retarded} form of \eqref{eq:green_function_expected_solution} 
       was obtained by these authors in \cite{aharonovich_2009}.
       
 \item in the physics (e.g. \cite{couranthilbert1989_2,GalTsov2002,Kazinski2002,kosyakov2007})  
       and mathematics (e.g. \cite{Gelfand1964_1,Kythe1996}) literature, 
       for fundamental solutions to the linear $N$-dimensional wave equation,
       many of which are a $t$-retarded form of \eqref{eq:green_function_expected_solution}, whereas the others are
       without specific retardation \cite{Gelfand1964_1}, and have a form closely related to \eqref{eq:green_function_expected_solution}.
\end{itemize}


Even though the previous methods have obtained different results, as displayed above, in this paper we shall show
that \emph{all of these methods} result in essentially the following form
\begin{align}
    \label{eq:green_function_expected_solution}
    g(x,\tau)
    & = 
        \dfrac{1}{2\pi^2}
        \lim\limits_{a \to 0^{+}}
        \dfrac{\partial}{\partial a}
        \dfrac{\theta(- x_{\mu} x^{\mu} - \tau^2 + a)}
              {\sqrt{- x_{\mu} x^{\mu} - \tau^2 + a} }
        \Big|_{a=0},
\end{align}
consistently with the solutions in the general mathematical literature \cite{Gelfand1964_1,Kythe1996}
(the form of \eqref{eq:green_function_expected_solution} implies a well defined regularization \cite{Gelfand1964_1}).

To the best of the our knowledge, however, 
\emph{explicit $\tau$-retarded solutions} could only be reproduced by methods of the type developed 
by Nozaki \cite{Nozaki_1964}, as was used in \cite{aharonovich_2009}.
For applications to physical problems such as that of self-interaction, we favor the $\tau$-retarded form.
This paper is primarily devoted to discussion of this form.
We discuss this point in the last section.

The remainder of the paper is organized as follows:
\begin{enumerate}
 \item  In section \ref{sec:land_method} we examine the method employed in ref. \cite{LandHor1991}.
        We shall term this method as \emph{the Klein-Gordon method}, since it essentially reproduces
        the Klein-Gordon propagator in the first 4D spacetime coordinates, and then integrates
        over $k_{5}$, essentially the Klein-Gordon \emph{mass} term. 
 
 \item  In section \ref{sec:oron_method} we examine the method of ref. \cite{OronHorwitz2000}, 
        which integrates over $k_{5}$ first, and then
        over the spacetime $k^{\mu}$ coordinates.

\end{enumerate}

\section{Klein-Gordon method}
\label{sec:land_method}
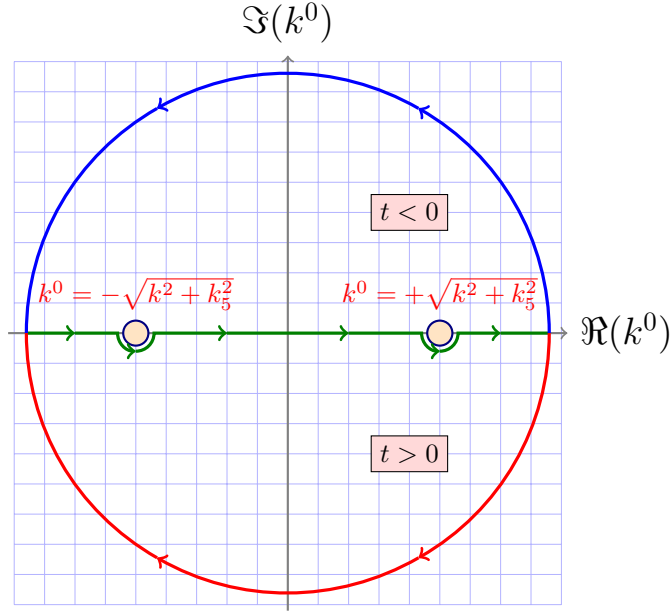
\begin{figure}
    \begin{center}
        \begin{tikzpicture}[scale=0.8]

            \draw[step=.5cm,color=blue!30,very thin] (-4.5, -4.5) grid (4.5,4.5);
            
            \draw [->,color=gray, thick] (-4.6,0) -- (4.6, 0)  node[right=1pt, color=black] {\Large $\Re (k^{0})$} coordinate (x axis);
            \draw [->,color=gray, thick] (0, -4.6) -- (0, 4.6) node[above=1pt, color=black] {\Large $\Im (k^{0})$} coordinate (y axis);
            \node [fill=Bisque1, draw=blue!50!black, circle,  thick] (negative pole) at (-2.5,0) {};
            \node [fill=Bisque1, draw=blue!50!black, circle,  thick] (positive pole) at ( 2.5,0) {};
            \node [red, above] at (negative pole.north) {$k^{0} = -\sqrt{k^2 + k_{5}^2}$};
            \node [red, above] at (positive pole.north) {$k^{0} = +\sqrt{k^2 + k_{5}^2}$};

            \draw[->, color=green!50!black, very thick] 
                    (-4.3, 0) -- (-3.5,0) ;
            \draw[->, color=green!50!black, very thick] 
                    (-3.5, 0) -- (-2.8,0) arc(-180:-90:0.3)  ;
            \draw[->, color=green!50!black, very thick] 
                   (-2.5,-0.3) arc(-90:0:0.3) -- (-1,0) ;
            \draw[->, color=green!50!black, very thick] 
                   (-1,0) -- (1,0);
            \draw[->, color=green!50!black, very thick] 
                   (1,0) -- (2.2,0) arc(-180:-90:0.3) ;
            \draw[->, color=green!50!black, very thick] 
                   (2.5,-0.3) arc(-90:0:0.3) -- (3.5,0) ;
            \draw[- , color=green!50!black, very thick] 
                    (3.5,0) -- (4.3,0);
            
            \draw[->, color=blue  , very thick] (4.3,0) arc(0:60:4.3) ;
            \draw[->, color=blue  , very thick] (60:4.3) arc(60:120:4.3) ;
            \draw[- , color=blue  , very thick] (120:4.3) arc(120:180:4.3) ;

            \draw[->, color=red   , very thick] (4.3,0) arc(0:-60:4.3) ;
            \draw[->, color=red   , very thick] (-60:4.3) arc(-60:-120:4.3) ;
            \draw[- , color=red   , very thick] (-120:4.3) arc(-120:-180:4.3) ;
            
            \node[fill=red!15,draw=black] at ( 2, 2) {$t < 0$} ;
            \node[fill=red!15,draw=black] at ( 2,-2) {$t > 0$} ;
        \end{tikzpicture}
        \caption{Contour integration for the Klein-Gordon Green function}
        \label{fig:klein_gordon_green_function_contour}
    \end{center}
\end{figure}

In the following, we discuss the analysis of Land and Horwitz \cite{LandHor1991}.
Starting from \eqref{eq:5d_green_function_fourier_transform}, we shall work with 3D spherical coordinates $(k,\theta,\phi)$.
After integrating over the spherical angles $(\theta,\phi)$, one integrates over $k_{0}$.
As the denominator has poles at $k_{0} = \pm \sqrt{\vb{k}^2 + k_{5}^2}$, the \emph{Principal Part} solution is taken,
using the contour given in figure \ref{fig:klein_gordon_green_function_contour}.

One can see that \eqref{eq:5d_green_function_fourier_transform} could be seen as an \emph{inverse Fourier transform}
in $m$ for the Klein-Gordon propagator:
\begin{align*}
    g(x,\tau)
    & = 
        \dfrac{1}{2\pi}
            \int_{-\infty}^{+\infty}
                e^{i m \tau}
                G_{KG}(x,m)
\end{align*}
where $G_{KG}$ is the Principal-Part Klein-Gordon GF with the well known form \cite{PhysRev.82.664}:
\begin{align}
    \label{eq:klein_gordon_propagator}
    G_{KG}(x,m)
    & = 
        - \dfrac{\delta(x_{\mu} x^{\mu})}{4\pi}
        + 
        \dfrac{m \theta(- x_{\mu} x^{\mu})}{4\pi} 
            J_{1}(m \sqrt{- x_{\mu} x^{\mu}})
\end{align}
We shall refer to \eqref{eq:klein_gordon_propagator} in due course.

Going back to \eqref{eq:5d_green_function_fourier_transform}, we find (after the integration over $\theta$ and $\phi$):
\begin{align*}
    g(x,\tau)
    & = 
        \dfrac{1}{(2\pi)^4}
        \dfrac{2}{r}
        \int_{-\infty}^{\infty}
            \ud k_{5}
        \int_{0}^{\infty}
            k \ud k
        \int_{-\infty}^{+\infty}
            \ud k_{0}
            \,
            \sin(kr)
            \dfrac{ e^{i(k_{5} \tau - k_{0}t)}   }
                  { k^2 - k_{0}^2 + k_{5}^2      }
    \\
    & = 
        -
        \dfrac{1}{(2\pi)^4}
        \dfrac{2}{r}
        \dfrac{\partial}{\partial r}
        \int_{-\infty}^{\infty}
            \ud k_{5}
        \int_{0}^{\infty}
            \ud k
        \int_{-\infty}^{+\infty}
            \ud k_{0}
            \,
            \cos(kr)
            \dfrac{ e^{i(k_{5} \tau - k_{0}t)}   }
                  { k^2 - k_{0}^2 + k_{5}^2      }
\end{align*}
where $k = |\vb{k}|$.

Now, the principal-part solution of the $k_{0}$ integral is
\begin{align*}
    \int_{-\infty}^{+\infty}
        \dfrac{e^{- ik_{0}t}}
              {k^2 + k_{5}^2 - k_{0}^2 }
        \ud k_{0}
    & = 
        i \pi \epsilon(-t)
        \left(
            a_{-1}(-)
            +
            a_{-1}(+)
        \right)
\end{align*}    
where
\begin{align*}
    a_{-1}(-)   
    & = 
        \left[
            \left( k_{0} + \sqrt{k^2 + k_{5}^2} \right)
            \times
            \dfrac{e^{- ik_{0}t}}
                  {k^2 + k_{5}^2 - k_{0}^2 }
        \right]_{k_{0} = - \sqrt{k^2 + k_{5}^2}}
    = 
            \dfrac{e^{+ i \sqrt{k^2 + k_{5}^2} t}}
                  { 2 \sqrt{k^2 + k_{5}^2}       }
    \\
    a_{-1}(+)   
    & = 
        \left[
            \left( k_{0} - \sqrt{k^2 + k_{5}^2} \right)
            \times
            \dfrac{e^{- ik_{0}t}}
                  {k^2 + k_{5}^2 - k_{0}^2 }
        \right]_{k_{0} = + \sqrt{k^2 + k_{5}^2}}
    = 
        -
            \dfrac{e^{- i \sqrt{k^2 + k_{5}^2} t}}
                  { 2 \sqrt{k^2 + k_{5}^2}       }
\end{align*}

Thus:
\begin{align*}
    g(x,\tau)
    & = 
        -
        \dfrac{i^2 \pi}{(2\pi)^4}
        \dfrac{2 \epsilon(-t)}{r}
        \dfrac{\partial}{\partial r}
        \int_{-\infty}^{\infty}
            \ud k_{5}
            e^{i k_{5} \tau}
        \int_{0}^{\infty}
            \ud k
            \,
            \cos(kr)
            \,
            \dfrac{     \sin \left( t \sqrt{k^2 + k_{5}^2} \right)  }
                  {     \sqrt{k^2 + k_{5}^2 }                       }
    \\
    & = 
        -
        \dfrac{1}{(2\pi)^3}
        \dfrac{\epsilon(t)}{r}
        \dfrac{1}{2}
        \dfrac{\partial}{\partial r}
        \int_{-\infty}^{\infty}
            \ud k_{5}
            e^{i k_{5} \tau}
        \int_{-\infty}^{\infty}
            \ud k
            \,
            \cos(kr)
            \,
            \dfrac{     \sin \left( t \sqrt{k^2 + k_{5}^2} \right)  }
                  {     \sqrt{k^2 + k_{5}^2 }                       }
\end{align*}
where we have extended the $k$ integration to the negative real axis as well, since the 
integrand is \emph{even in $k$}. Further progress is made by substituting 
$k(\beta) = |k_{5}| \sinh( \beta) $\footnote{$|k_{5}|$ ensures the bounds on $\beta$ are invariant under the sign of $k_{5}$.}:
\begin{align*}
    g(x,\tau)
    & = 
        -
        \dfrac{1}{(2\pi)^3}
        \dfrac{\epsilon(t)}{r}
        \dfrac{1}{2}
        \dfrac{\partial}{\partial r}
        \int_{-\infty}^{\infty}
            \ud k_{5}
            e^{i k_{5} \tau}
        \int_{-\infty}^{\infty}
            (|k_{5}| \cosh(\beta) \ud \beta) 
            \times
            \,
            \cos(r |k_{5}| \sinh(\beta) )
            \,
            \dfrac{     \sin \left( t |k_{5}| \cosh(\beta) \right)  }
                  {     |k_{5}| \cosh(\beta)                        }
    \\
    & = 
        -
        \dfrac{1}{(2\pi)^3}
        \dfrac{\epsilon(t)}{r}
        \dfrac{1}{2}
        \dfrac{\partial}{\partial r}
        \int_{-\infty}^{\infty}
            \ud k_{5}
            e^{i k_{5} \tau}
        \int_{-\infty}^{\infty}
            \ud \beta
            \times
            \,
            \cos(r |k_{5}| \sinh(\beta) )
            \,
            \sin \left( t |k_{5}| \cosh(\beta) \right) 
    \\
    & = 
        -
        \dfrac{1}{(2\pi)^3}
        \dfrac{\epsilon(t)}{r}
        \dfrac{1}{2}
        \dfrac{\partial}{\partial r}
        \int_{-\infty}^{\infty}
            \ud k_{5}
            e^{i k_{5} \tau}
        \int_{-\infty}^{\infty}
            \ud \beta
    \\
    & \qquad \qquad 
        \times
            \dfrac{1}{2}
            \left[
                \sin
                    \left(
                        |k_{5}|
                        (
                            r \sinh(\beta)
                            +
                            t \cosh(\beta)
                        )
                    \right)
                -
                \sin
                    \left(
                        |k_{5}|
                        (
                            r \sinh(\beta)
                            -
                            t \cosh(\beta)
                        )
                    \right)
            \right]
\end{align*}

If $|t| < r$, we can substitute:
\begin{align*}
    t & = \rho \sinh(\alpha)    
    &   
    r & = \rho \cosh(\alpha)
    &
    \rho^2 & = r^2 - t^2
\end{align*}
and thus:
\begin{align*}
    g(x,\tau)
    & = 
        -
        \dfrac{1}{(2\pi)^3}
        \dfrac{2\epsilon(t)}{r}
        \dfrac{\partial}{\partial r}
        \int_{-\infty}^{\infty}
            \ud k_{5}
            e^{i k_{5} \tau}
        \int_{0}^{\infty}
            \,
            \ud \beta
    \\
    & \qquad \qquad 
        \times
            \dfrac{\theta(r^2 - t^2)}{2}
            \left[
                \sin
                    \left(
                        |k_{5}| \rho \sinh(\beta + \alpha)
                    \right)
                -
                \sin
                    \left(
                        |k_{5}| \rho \sinh(- \beta + \alpha)
                    \right)
            \right]
    \\
    & = 
        0
\end{align*}
The result for $r^2 > t^2$ has the integrand $\sin(|k_{5}| \rho \sinh(\alpha \pm \beta))$ which is 
\emph{odd} around the center $\beta \mp \alpha = 0$, and since the bounds are even at $\pm \infty$,
we obtain the null result.

On the other hand, when $|t| > r$ we find:
\begin{align*}
    t & = \epsilon(t) \rho \cosh(\alpha)    
    &   
    r & = \rho \sinh(\alpha)
    &
    \rho^2 & = t^2 - r^2
\end{align*}

And thus:
\begin{align*}
    t \cosh(\beta) + r \sinh(\beta) 
    & = 
        \epsilon(t) \rho \cosh(\alpha) \cosh(\beta)
        +
        \rho \sinh(\alpha) \sinh(\beta)
    \\
    & = 
        \epsilon(t) \rho \cosh(\alpha + \epsilon(t) \beta)
    \\
    r \sinh(\beta) - t \cosh(\beta) 
    & = 
        - (t \cosh(\beta) - r \sinh(\beta))
    \\
    & =  
        \epsilon(-t) \rho \cosh(\alpha + \epsilon(-t) \beta )
\end{align*}

Substituting back in $g(x,\tau)$ we find:
\begin{align*}
    g(x,\tau)
    & = 
        -
        \dfrac{1}{(2\pi)^3}
        \dfrac{2\epsilon(t)}{r}
        \dfrac{\partial}{\partial r}
        \int_{-\infty}^{\infty}
            \ud k_{5}
            e^{i k_{5} \tau}
        \int_{-\infty}^{\infty}
            \,
            \ud \beta
    \\
    & \qquad \qquad 
        \times
            \dfrac{\theta(t^2 - r^2)}{2}
            \left[
                \sin
                \left(
                    \epsilon(t) |k_{5}| \rho \cosh(\alpha + \epsilon(t) \beta)
                \right)
                -
                \sin
                \left(
                    \epsilon(-t) |k_{5}| \rho \cosh(\alpha + \epsilon(-t) \beta) 
                \right)
            \right]
    \\
    & = 
        -
        \dfrac{1}{(2\pi)^3}
        \dfrac{1}{r}
        \dfrac{\partial}{\partial r}
        \int_{-\infty}^{\infty}
            \ud k_{5}
            e^{i k_{5} \tau}
        \int_{-\infty}^{\infty}
            \,
            \ud \beta
            \,
            \times
            \theta(t^2 - r^2)
            \times 
            \sin
            \left(
                |k_{5}| \rho \cosh(\beta)
            \right)
\end{align*}
Substituting $u = \cosh(\beta)$ we find:
\begin{align}
    g(x,\tau)
    & = 
        -
        \dfrac{1}{(2\pi)^3}
        \dfrac{1}{r}
        \dfrac{\partial}{\partial r}
        \int_{-\infty}^{\infty}
            \ud k_{5}
            e^{i k_{5} \tau}
        \times
        2 
        \times
        \int_{1}^{\infty}
            \,
            \dfrac{\ud u}{\sqrt{u^2 - 1}}
            \,
            \times
            \theta(t^2 - r^2)
            \times 
            \sin
            \left(
                |k_{5}| \rho u
            \right)
    \nonumber \\
    \label{eq:martin_green_function_prepamture}
    & = 
        -
        \dfrac{1}{(2\pi)^3}
        \dfrac{2}{r}
        \dfrac{\partial}{\partial r}
        \times
        \theta(t^2 - r^2)
        \int_{1}^{\infty}
            \,
            \dfrac{\ud u}{\sqrt{u^2 - 1}}
        \int_{-\infty}^{\infty}
            \ud k_{5}
            e^{i k_{5} \tau}
            \,
            \sin
            \left(
                |k_{5}| \rho u
            \right)
\end{align}

As the $k_{5}$ integration picks up only the \emph{even} part, we can rewrite it as follows:
\begin{align*}
    g(x,\tau)
    & = 
        -
        \dfrac{1}{(2\pi)^3}
        \dfrac{2}{r}
        \dfrac{\partial}{\partial r}
        \times
        \theta(t^2 - r^2)
        \int_{1}^{\infty}
            \,
            \dfrac{\ud u}{\sqrt{u^2 - 1}}
            \times
            2 
            \times
        \int_{0}^{\infty}
            \ud k_{5}
            \cos(k_{5} \tau)
            \,
            \sin
            \left(
                k_{5} \rho u
            \right)
    \\
    & = 
        -
        \dfrac{1}{(2\pi)^3}
        \dfrac{2}{r}
        \dfrac{\partial}{\partial r}
        \times
        \theta(t^2 - r^2)
        \int_{1}^{\infty}
            \,
            \dfrac{\ud u}{\sqrt{u^2 - 1}}
    \\
    & \qquad \qquad
        \times
        2 
        \times
        \int_{0}^{\infty}
            \ud k_{5}
            \dfrac{1}{4i}
            \left[
                e^{i k_{5} (\tau + \rho u)}
                - 
                e^{i k_{5} (\tau - \rho u)}
                + 
                e^{i k_{5} (- \tau + \rho u)}
                - 
                e^{- i k_{5} (\tau + \rho u)}
            \right]
\end{align*}

Since\footnote{E.g., in \cite{Gelfand1964_1}.}
\begin{align}
    \int_{0}^{\infty} 
        \ud k_{5}
        e^{i k_{5} (\tau + \rho u)}
    & = 
        + \pi \delta(\tau + \rho u) - P\dfrac{1}{i(\tau + \rho u)}
\end{align}
we find:
\begin{align}
    g(x,\tau)
    & = 
        -
        \dfrac{1}{(2\pi)^3}
        \dfrac{2}{r}
        \dfrac{\partial}{\partial r}
        \times
        \theta(t^2 - r^2)
        \int_{1}^{\infty}
            \,
            \dfrac{\ud u}{\sqrt{u^2 - 1}}
        \nonumber
    \\
    & \qquad \qquad 
        \times
        \dfrac{1}{2i}
        \left[
            + 
            \pi
            \left(
                \delta(\tau + \rho u)
                - 
                \delta(\tau - \rho u)
                +
                \delta(- \tau + \rho u)
                - 
                \delta(- \tau - \rho u)
            \right)
        \right.
        \nonumber
    \\
    & \qquad \qquad 
        \left.
            -
            P \dfrac{1}{i(\tau + \rho u)}
            + 
            P \dfrac{1}{i(\tau - \rho u)}
            -
            P \dfrac{1}{i(-\tau + \rho u)}
            + 
            P \dfrac{1}{i(-\tau - \rho u)}
        \right]
        \nonumber
    \\
    \label{eq:g_x_tau_before_u_integration}
    & =
        -
        \dfrac{1}{(2\pi)^3}
        \dfrac{2}{r}
        \dfrac{\partial}{\partial r}
        \times
        \theta(t^2 - r^2)
        \int_{1}^{\infty}
            \,
            \dfrac{\ud u}{\sqrt{u^2 - 1}}
            \times
            P
            \left[
                \dfrac{1}{\rho u + \tau}
                +
                \dfrac{1}{\rho u - \tau}
            \right]
\end{align}

Let us inspect an integral of the form:
\begin{align*}
    I(a,b)
    & = 
        \int_{1}^{\infty}
            \dfrac{\ud x}{\sqrt{x^2 - 1}}
            \dfrac{1}{ax + b}
\end{align*}
Substituting $x(\alpha) = \cosh(\alpha), \ud x(\alpha) = \sinh(\alpha) \ud \alpha$ we find:
\begin{align*}
    I(a,b)
    & = 
        \int_{0}^{\infty}
            \dfrac{\sinh(\alpha) \ud \alpha }{\sinh \alpha}
            \dfrac{1}{a \cosh(\alpha) + b}
    =
        \int_{0}^{\infty}
            \dfrac{\ud \alpha }{a \cosh(\alpha) + b}
    =
        \dfrac{1}{2}
        \times
        \int_{-\infty}^{\infty}
            \dfrac{\ud \alpha }{a \cosh(\alpha) + b}
\end{align*}
where we have utilized the evenness of $\cosh(\alpha)$ around $\alpha=0$.

After a further substitution of $u(\alpha) = e^{\alpha}, \ud \alpha(u) = \ud u / u$, we find:
\begin{align*}
    I(a,b)
    & = 
        \dfrac{1}{2}
        \int_{0}^{\infty}
            \dfrac{\ud u / u }{a \frac{1}{2}(u + 1/u) + b}
    =
        \dfrac{1}{a}
        \int_{0}^{\infty}
            \dfrac{\ud u }{u^2 + 2ub/a + 1}
\end{align*}
The roots of the denominator are:
\begin{align*}
    u_{1,2} & = - \dfrac{b}{a} \pm \sqrt{ \dfrac{b^2}{a^2} - 1} 
             = \dfrac{1}{a}
               \left[
                   -b \pm \sqrt{b^2 - a^2}
               \right]
\end{align*}
If $b^2 < a^2$, then we can rewrite the denominator as:
\begin{align*}
    u^2 + 2u\dfrac{b}{a} + 1 
    & = 
        \left( u + \dfrac{b}{a} \right)^{2} + 1 - \dfrac{b^2}{a^2}
\end{align*}

Therefore, if $a^2 < b^2$, we have:
\begin{align*}
    I(a,b)
    & = 
        \dfrac{1}{a}
        \int_{0}^{\infty}
            \dfrac{\ud u }{(u - u_{1}) (u - u_{2})}
    =
        \dfrac{1}{a}
        \int_{0}^{\infty}
        \dfrac{\ud u}{u_{1} - u_{2}}
        \left[
            \dfrac{1}{u - u_{1}}
            - 
            \dfrac{1}{u - u_{2}}
        \right]
    \\
    & =
        \dfrac{1}{a(u_{1} - u_{2})}
        \ln 
        \Big|
            \dfrac{u - u_{1}}
                  {u - u_{2}}
        \Big|_{0}^{\infty}
    =
        \dfrac{1}{a(u_{1} - u_{2})}
        \ln 
        \Big|
            0
            - 
            \dfrac{u_{1}}{u_{2}}
        \Big|
    \\
    & =
        -
        \dfrac{1}{2 a \sqrt{b^2 - a^2} }
        \ln 
        \Big|
            \dfrac{b + \sqrt{b^2 - a^2}}
                  {b - \sqrt{b^2 - a^2} }
        \Big|
\end{align*}
where we have taken the principal part of the integration, 
as in \eqref{eq:g_x_tau_before_u_integration}.

However, in our case, eq.  \eqref{eq:g_x_tau_before_u_integration}, 
we have (with principal part)
$I(a,b) + I(a,-b)$, but $I(a,b) = - I(a,-b)$.
Therefore, when $b^2 > a^2$, which corresponds to $\tau^2 > \rho^2$, $g(x,\tau)$ vanishes..

In the other case where $b^2 < a^2$, we find:
\begin{align*}
    I(a,b)
    & = 
        \dfrac{1}{a}
        \int_{0}^{\infty}
            \dfrac{\ud u }{u^2 + 2ub/a + 1}
    =
        \dfrac{1}{a}
        \int_{0}^{\infty}
            \dfrac{\ud u }
                  {(u + b/a)^2 + D^2}
\end{align*}
where we have put $D^2 = 1 - b^2 / a^2$.

Substituting $v = (u + b/a)/D$, we find:
\begin{align*}
    I(a,b)
    & = 
        \dfrac{1}{a}
        \int_{b/aD}^{\infty}
            \dfrac{D \ud v }
                  {v^2 D^2 + D^2}
    \\
    & = 
        \dfrac{1}{a}
        \int_{b/aD}^{\infty}
            \dfrac{D \ud v }
                  {v^2 D^2 + D^2}
    =
        \dfrac{1}{aD}
        \int_{b/aD}^{\infty}
            \dfrac{\ud v }
                  {v^2 + 1}
    \\
    & = 
        \dfrac{1}{aD}
        \tan^{-1}(v) \Big|_{b/aD}^{\infty}
    =
        \dfrac{1}{aD}
        \left[
            \dfrac{\pi}{2}
            - 
            \tan^{-1} \left( \dfrac{b}{aD}\right)
        \right]
    \\
    & = 
        \dfrac{1}{a \sqrt{1 - b^2 / a^2}}
        \left[
            \dfrac{\pi}{2}
            - 
            \tan^{-1} \left( \dfrac{b}{a\sqrt{1 - b^2 / a^2}}\right)
        \right]
    =
        \dfrac{1}{\sqrt{a^2 - b^2 }}
        \left[
            \dfrac{\pi}{2}
            - 
            \tan^{-1} \left( \dfrac{b}{\sqrt{a^2 - b^2}}\right)
        \right]
\end{align*}
Now:
\begin{align*}
    I(a,b)
    & = 
        \dfrac{1}{\sqrt{a^2 - b^2 }}
        \left[
            \dfrac{\pi}{2}
            - 
            \tan^{-1} \left( \dfrac{b}{\sqrt{a^2 - b^2}}\right)
        \right]
    \\
    I(a,-b)
    & = 
        \dfrac{1}{\sqrt{a^2 - b^2 }}
        \left[
            \dfrac{\pi}{2}
            + 
            \tan^{-1} \left( \dfrac{b}{\sqrt{a^2 - b^2}}\right)
        \right]
\end{align*}
And thus:
\begin{align*}
    I(a,b) + I(a,-b)
    & = 
        \dfrac{\pi}{\sqrt{a^2 - b^2 }}
\end{align*}

Thus, we are left with the solution
\begin{align*}
    g(x,\tau)
    & = 
        -
        \dfrac{1}{(2\pi)^3}
        \dfrac{2}{r}
        \dfrac{\partial}{\partial r}
        \times
        \theta(t^2 - r^2)
        \times
        \dfrac{\pi \theta(\rho^2 - \tau^2)}{\sqrt{\rho^2 - \tau^2}}
    \\
    & = 
        \dfrac{1}{4\pi^2}
        \dfrac{1}{r}
        \dfrac{\partial}{\partial r}
        \dfrac{\theta(t^2 - r^2) \times \theta(t^2 - r^2 - \tau^2)}{\sqrt{t^2 - r^2 - \tau^2}}
\end{align*}

Now, clearly, $\theta(t^2 - r^2) \times \theta(t^2 - r^2 - \tau^2) = \theta(t^2 - r^2 - \tau^2)$.
Moreover, writing 
\begin{align*}
    \dfrac{1}{r}
    \dfrac{\partial}{\partial r}
    & = 
        \dfrac{1}{r} \dfrac{\partial r^2}{\partial r} \dfrac{\partial}{\partial r^2}
    =
        \dfrac{2r}{r}  \dfrac{\partial}{\partial r^2}
    =
        2 
        \dfrac{\partial}{\partial r^2}
\end{align*}
we find:
\begin{align*}
    g(x,\tau)
    & = 
        \dfrac{1}{2\pi^2}
        \dfrac{\partial}{\partial r^2}
        \dfrac{\theta(t^2 - r^2 - \tau^2)}{\sqrt{t^2 - r^2 - \tau^2}}
\end{align*}
and this is the GF expected, which differs from the one found by Land and Horwitz \cite{LandHor1991}, i.e.,
\begin{align}
    \label{eq:martin_green_function}
    G_{P}(x,\tau)
    & = 
        - 
        \dfrac{1}{4\pi} \delta(t^2 - r^2) \delta(\tau)
        - 
        \dfrac{1}{4\pi^2} 
        \dfrac{\partial}{\partial r^2}
        \dfrac{\theta(t^2 - r^2 - \tau^2)}{\sqrt{t^2 - r^2 - \tau^2}}
\end{align}

The reason is due to carrying out the derivative of $\theta(t^2 - r^2)$  in
\eqref{eq:martin_green_function_prepamture}, \emph{before} performing the integration.
The integration results in a $\theta(t^2 - r^2 - \tau^2)$ factor that \emph{subsumes} the $\theta(t^2 - r^2)$ prefactor.
And therefore, the first term in \eqref{eq:martin_green_function} \emph{should not appear}.
Taking this derivative \emph{prematurely}, results in the following construction, which 
leads directly to \eqref{eq:martin_green_function}:
\begin{align*}
    g(x,\tau)
    & = 
        -
        \dfrac{1}{(2\pi)^3}
        \dfrac{4}{r}
        \dfrac{\partial}{\partial r}
        \times
        \theta(t^2 - r^2)
        \int_{0}^{\infty}
            \ud k_{5}
            \cos(k_{5} \tau)
            \,
        \int_{1}^{\infty}
            \dfrac{\ud u}{\sqrt{u^2 - 1}}
            \,
            \sin
            \left(
                k_{5} \rho u
            \right)
    \\
    & = 
        -
        \dfrac{1}{(2\pi)^3}
        \dfrac{(-2r)}{r}
        \delta(t^2 - r^2)
        \int_{0}^{\infty}
            \ud k_{5}
            \cos(k_{5} \tau)
            \,
        \overbrace
        {
            \int_{1}^{\infty}
                \,
                \dfrac{\ud u}{\sqrt{u^2 - 1}}
                \,
                \sin
                \left(
                    |k_{5}| \rho u
                \right)
        }^{\pi J_{0}(k_{5} \rho) / 2}
    \\
    & \qquad
        -
        \dfrac{1}{(2\pi)^3}
        \dfrac{2}{r}
        \theta(t^2 - r^2)
        \times
        \dfrac{\partial}{\partial r}
        \int_{1}^{\infty}
            \,
            \dfrac{\ud u}{\sqrt{u^2 - 1}}
        \int_{0}^{\infty}
            \ud k_{5}
            \cos(k_{5} \tau)
            \,
            \sin
            \left(
                |k_{5}| \rho u
            \right)
\end{align*}
We can rewrite it as:
\begin{align*}
    g(x,\tau)
    & = 
        \dfrac{1}{2\pi}
        \int_{-\infty}^{\infty}
            \ud k_{5}
            \,
            e^{i k_{5} \tau}
            G_{KG}(x, k_{5})
\end{align*}
where $G_{KG}(x,k_{5})$ is given in \eqref{eq:klein_gordon_propagator}.

Moving on with the integration, we immediately find:
\begin{align*}
    g(x,\tau)
    & = 
        \dfrac{1}{8 \pi^2 }
        \delta(t^2 - r^2)
        \int_{0}^{\infty}
            \ud k_{5}
            \cos(k_{5} \tau)
            \,
            J_{0}(k_{5} \rho)
        +
        \dfrac{\theta(t^2 - r^2)}{2\pi^2}
        \dfrac{\partial}{\partial r^2}
        \dfrac{\theta(t^2 - r^2 - \tau^2)}{\sqrt{t^2 - r^2 - \tau^2}}
    \\
    & = 
        \dfrac{1}{4 \pi^2 }
        \delta(t^2 - r^2)
        \dfrac{1}{2}
        \int_{-\infty}^{\infty}
            \ud k_{5}
            \cos(k_{5} \tau)
            \times 
            1
        +
        \dfrac{\theta(t^2 - r^2)}{2\pi^2}
        \dfrac{\partial}{\partial r^2}
        \dfrac{\theta(t^2 - r^2 - \tau^2)}{\sqrt{t^2 - r^2 - \tau^2}}
    \\
    & = 
        \dfrac{1}{4 \pi }
        \delta(t^2 - r^2)
        \delta(\tau)
        +
        \dfrac{\theta(t^2 - r^2)}{2\pi^2}
        \dfrac{\partial}{\partial r^2}
        \dfrac{\theta(t^2 - r^2 - \tau^2)}{\sqrt{t^2 - r^2 - \tau^2}}
\end{align*}

We see that extra term $\delta(t^2 - r^2) \delta(\tau)$ arises from the condition
on the Klein-Gordon Green-function \eqref{eq:klein_gordon_propagator} which results in a contribution
\emph{on} the 4D light-cone $x_{\mu} x^{\mu}$.
However, we note that integrating $G_{KG}(x,m)$ \emph{directly} with respect to $m$ 
\emph{does not reproduce} the 5D GF as in \eqref{eq:green_function_expected_solution}.
To see this we first show that it is indeed \emph{expected} that such an integration would produce \eqref{eq:green_function_expected_solution},
as $G_{KG}(x,m)$ is essentially the GF of the Klein-Gordon equation
\begin{align}
    \label{eq:klein_gordon_equation}
    (\partial_{\mu} \partial^{\mu} - m^2)G_{KG}(x,m) & = \delta^4(x)
\end{align}
and therefore, integrating over $m$
\begin{align}
    \int_{-\infty}^{+\infty} e^{i m \tau} 
                (\partial_{\mu} \partial^{\mu} - m^2)G_{KG}(x,m) 
                \ud m
    & = 
        \delta^{4}(x)
        \,
        \int_{-\infty}^{+\infty} e^{i m \tau} 
                \ud m
    \\
    \int_{-\infty}^{+\infty} e^{i m \tau} 
                (\partial_{\mu} \partial^{\mu} + \partial_{\tau^2})G_{KG}(x,m) 
                \ud m
    & = 
        2\pi
        \delta^{4}(x)
        \,
        \delta(\tau)
\end{align}
and therefore 
\begin{align}
    \label{eq:green_function_from_klein_gordon_propagator}
    g(x,\tau)
    & = 
        \dfrac{1}{2\pi }
        \int_{-\infty}^{+\infty}
            e^{i m \tau}
            G_{KG}(x,m)
            \ud m
\end{align}
as expected. Nevertheless, taking $G_{KG}(x,m)$ directly as in \eqref{eq:klein_gordon_propagator}
would not produce \eqref{eq:green_function_expected_solution}.
However, one can rewrite \eqref{eq:klein_gordon_propagator} in such a way that results  in 
\eqref{eq:green_function_from_klein_gordon_propagator} correctly, as follows:
\begin{align}
    \label{eq:klein_gordon_propagator_correct_version}
    G_{KG}(x,m)
    & = 
        \dfrac{\partial}{\partial r^2}
        \Big[
            \dfrac{m \theta(-x_{\mu} x^{\mu})  }
                  {4 \pi                       }
            J_{0}(m \sqrt{- x_{\mu} x^{\mu}})
        \Big]
\end{align}
Therefore, under the integral in \eqref{eq:green_function_from_klein_gordon_propagator}, the
$\theta(-x_{\mu} x^{\mu})$ in eq. \eqref{eq:klein_gordon_propagator_correct_version} would be subsumed 
by $\theta(-x_{\mu} x^{\mu} - \tau^2)$, in accordance with \eqref{eq:green_function_expected_solution},
and therefore, would eliminate the $\delta(x_{\mu} x^{\mu}) \delta(\tau)$.


\section{Integration over $k_{5}$ first}
\label{sec:oron_method}

In this method, Oron and Horwitz \cite{OronHorwitz2000} split \eqref{eq:5d_green_function_fourier_transform} into $2$ regions
in $(k, k_{0})$ space, the timelike region $k_{\mu} k^{\mu} < 0$ and the spacelike region $k_{\mu} k^{\mu} > 0$.
We shall reexamine this calculation here and find that correcting a sign error in \cite{OronHorwitz2000}, 
the result \eqref{eq:green_function_expected_solution}
emerges from this method as well.

Thus, one finds 
\begin{align*}
    g(x,\tau) & = g_{1}(x,\tau) + g_{2}(x,\tau)
    \\
    g_{1}(x,\tau)
    & = 
        \dfrac{1}{(2\pi)^4}
        \dfrac{2}{r}
        \int_{-\infty}^{\infty}
            \ud k_{5}
        \int_{0}^{\infty}
            k \ud k
        \int_{-\infty}^{+\infty}
            \ud k_{0}
            \,
            \theta(k^2 - k_{0}^2)
            \sin(kr)
            \dfrac{ e^{i(k_{5} \tau - k_{0}t)}   }
                  { k^2 - k_{0}^2 + k_{5}^2      }
    \\
    g_{2}(x,\tau)
    & = 
        \dfrac{1}{(2\pi)^4}
        \dfrac{2}{r}
        \int_{-\infty}^{\infty}
            \ud k_{5}
        \int_{0}^{\infty}
            k \ud k
        \int_{-\infty}^{+\infty}
            \ud k_{0}
            \,
            \theta(- k^2 + k_{0}^2)
            \sin(kr)
            \dfrac{ e^{i(k_{5} \tau - k_{0}t)}   }
                  { k^2 - k_{0}^2 + k_{5}^2      }
\end{align*}

Then each of the functions can be contour integrated over $k_{5}$. Clearly, 
in $g_{1}(x,\tau)$ the integral is well defined as the poles are in the complex plane
$k_{5} = \pm i \sqrt{k^2 - k_{0}^2}$, whereas in $g_{2}(x,\tau)$, the Principal Part
is taken over $k_{5} = \pm \sqrt{k_{0}^2 - k^2}$.

The result is
\begin{align*}
    g_{1}(x,\tau)
    & =
        \dfrac{1}{(2\pi)^3 r}
        \int_{0}^{\infty}
            \ud l
        \int_{-\infty}^{+\infty}
            \ud \alpha
            \,\,
            l \cosh(\alpha)
            \sin(l r \cosh(\alpha))
            \cos(l t \sinh(\alpha))
            e^{- l |\tau|}
    \\
    g_{2}(x,\tau)
    & = 
        -
        \dfrac{1}
              {(2\pi)^3 r      }
        \int_{0}^{\infty}
            \ud l
        \int_{-\infty}^{+\infty}
            \ud \alpha
            \,\,
            l \sinh(\alpha)
            \sin(l r \sinh(\alpha))
            \cos(l t \cosh(\alpha))
            \sin(l |\tau|)
\end{align*}
where $l = \sqrt{ \pm (k^2 - k_{0}^2)}$ and $\alpha$ is the corresponding hyperbolic angle.

In both cases we can simplify by absorbing $l \cosh(\alpha)$ or $l \sinh(\alpha)$ as follows:
\begin{align*}
    g_{1}(x,\tau)
    & =
        -
        \dfrac{1}{(2\pi)^3 r}
        \dfrac{\partial}{\partial r}
        \int_{0}^{\infty}
            \ud l
        \int_{-\infty}^{+\infty}
            \ud \alpha
            \,\,
            \cos(l r \cosh(\alpha))
            \cos(l t \sinh(\alpha))
            e^{- l |\tau|}
    \\
    g_{2}(x,\tau)
    & = 
        \dfrac{1}
              {(2\pi)^3 r      }
        \dfrac{\partial}{\partial r}
        \int_{0}^{\infty}
            \ud l
        \int_{-\infty}^{+\infty}
            \ud \alpha
            \,\,
            \cos(l r \sinh(\alpha))
            \cos(l t \cosh(\alpha))
            \sin(l |\tau|)
\end{align*}
Expanding the $\cos(\ldots)$ terms:
\begin{align}
    \label{eq:starting_point_g1_and_g2}
    g_{1}(x,\tau)
    & =
        -
        \dfrac{1}{2(2\pi)^3 r}
        \dfrac{\partial}{\partial r}
        \int_{0}^{\infty}
            \ud l
        \int_{-\infty}^{+\infty}
            \ud \alpha
            \,\,
    \nonumber \\
    & \qquad \qquad 
            \times
            \Big(
                \cos \left( l (r \cosh(\alpha) + t \sinh(\alpha) \right)
                +
                \cos \left( l (r \cosh(\alpha) - t \sinh(\alpha) \right)
            \Big)
            e^{- l |\tau|}
    \\
    g_{2}(x,\tau)
    & = 
        \dfrac{1}
              {2(2\pi)^3 r     }
        \dfrac{\partial}{\partial r}
        \int_{0}^{\infty}
            \ud l
        \int_{-\infty}^{+\infty}
            \ud \alpha
            \,\,
    \nonumber \\
    & \qquad \qquad 
            \times
            \Big(
                \cos \left( l (r \sinh(\alpha) + t \cosh(\alpha) \right)
                +
                \cos \left( l (r \sinh(\alpha) - t \cosh(\alpha) \right)
            \Big)
            \sin(l |\tau|)
    \nonumber
\end{align}
For $r > |t|$ we can write $r = \rho \cosh(\beta)$ and $t = \rho \sinh(\beta)$ to find:
\begin{align*}
    g_{1}(x,\tau)
    & =
        -
        \dfrac{1}{2(2\pi)^3 r}
        \dfrac{\partial}{\partial r}
        \int_{0}^{\infty}
            \ud l
        \int_{-\infty}^{+\infty}
            \ud \alpha
            \,\,
    \\
    & \qquad \qquad 
            \times
            \Big(
                \cos \left( l \rho \cosh(\alpha + \beta) \right)
                +
                \cos \left( l \rho \cosh(\alpha - \beta) \right)
            \Big)
            e^{- l |\tau|}
    \\
    g_{2}(x,\tau)
    & = 
        \dfrac{1}
              {2(2\pi)^3 r     }
        \dfrac{\partial}{\partial r}
        \int_{0}^{\infty}
            \ud l
        \int_{-\infty}^{+\infty}
            \ud \alpha
            \,\,
    \\
    & \qquad \qquad 
            \times
            \Big(
                \cos \left( l \rho \sinh(\alpha + \beta) \right)
                +
                \cos \left( l \rho \sinh(\alpha - \beta) \right)
            \Big)
            \sin(l |\tau|)
\end{align*}
Clearly, the symmetry of the integration bounds on $\alpha$ indicate symmetry of the two summed terms.
\begin{align*}
    g_{1}(x,\tau)
    & =
        -
        \dfrac{1}{(2\pi)^3 r}
        \dfrac{\partial}{\partial r}
        \int_{0}^{\infty}
            \ud l
        \int_{-\infty}^{+\infty}
            \ud \alpha
            \,\,
            \times
            \cos \left( l \rho \cosh(\alpha) \right)
            e^{- l |\tau|}
    \\
    g_{2}(x,\tau)
    & = 
        \dfrac{1}
              {(2\pi)^3 r      }
        \dfrac{\partial}{\partial r}
        \int_{0}^{\infty}
            \ud l
        \int_{-\infty}^{+\infty}
            \ud \alpha
            \,\,
            \times
            \cos \left( l \rho \sinh(\alpha ) \right)
            \sin(l |\tau|)
\end{align*}
Performing the $l$ integration first, one obtains:
\begin{align*}
    g_{1}(x,\tau)
    & =
        -
        \dfrac{1}{(2\pi)^3 r}
        \dfrac{\partial}{\partial r}
        \int_{-\infty}^{+\infty}
            \ud \alpha
        \int_{0}^{\infty}
            \ud l
            \,\,
            \times
            \cos \left( l \rho \cosh(\alpha) \right)
            e^{- l |\tau|}
    \\
    & = 
        -
        \dfrac{1}{(2\pi)^3 r}
        \dfrac{\partial}{\partial r}
        \int_{-\infty}^{+\infty}
            \ud \alpha
            \dfrac{1}{2}
            \left[
                \dfrac{1}{|\tau| - i \rho \cosh(\alpha)}
                +
                \dfrac{1}{|\tau| + i \rho \cosh(\alpha)}
            \right]
    \\
    & = 
        \dfrac{1}{(2\pi)^3 r}
        \dfrac{\partial}{\partial r}
        \int_{-\infty}^{+\infty}
            \ud \alpha
            \dfrac{1}{2i}
            \left[
                \dfrac{1}{\rho \cosh(\alpha) - i|\tau| }
                -
                \dfrac{1}{\rho \cosh(\alpha) + i|\tau|}
            \right]
    \\
    g_{2}(x,\tau)
    & = 
        \dfrac{1}
              {(2\pi)^3 r      }
        \dfrac{\partial}{\partial r}
        \int_{-\infty}^{+\infty}
            \ud \alpha
        \int_{0}^{\infty}
            \ud l
            \,\,
            \times
            \cos \left( l \rho \sinh(\alpha ) \right)
            \sin(l |\tau|)
    \\
    & = 
        \dfrac{1}
              {(2\pi)^3 r      }
        \dfrac{\partial}{\partial r}
        \int_{-\infty}^{+\infty}
            \ud \alpha
            \dfrac{1}{2}
            \left[
                \dfrac{1}{\rho \sinh(\alpha) + |\tau|}
                -
                \dfrac{1}{\rho \sinh(\alpha) - |\tau|}
            \right]
\end{align*}

Integrating $g_{1}(x,\tau)$ over $\alpha$ results in
\begin{align*}
    g_{1}(x,\tau) 
    & = 
        \dfrac{1}{2i}
        \dfrac{1}{(2\pi)^3 r}
        \dfrac{\partial}{\partial r}
        2 
        \times
        \dfrac{1}{\sqrt{(i|\tau|)^2 - \rho^2}}
        \ln
        \left(
            \dfrac{   i|\tau| + \sqrt{(i|\tau|)^2 - \rho^2}   }
                  { - i|\tau| + \sqrt{(i|\tau|)^2 - \rho^2}   }
        \right)
    \\
    & = 
        \dfrac{1}{i}
        \dfrac{1}{(2\pi)^3 r}
        \dfrac{\partial}{\partial r}
        \dfrac{1}{\sqrt{- \tau^2  -\rho^2}}
        \ln
        \left(
            \dfrac{   i|\tau| + \sqrt{- \tau^2  -\rho^2}    }
                  { - i|\tau| + \sqrt{- \tau^2  -\rho^2}    }
        \right)
\end{align*}
and since $\rho^2 = r^2 - t^2 > 0$, we have
\begin{align*}
    \sqrt{- \tau^2 - \rho^2} & =
        \sqrt{(-1)(\rho^2 + \tau^2)} 
    =
        i \sqrt{\rho^2 + \tau^2} 
\end{align*}
and thus:
\begin{align*}
    g_{1}(x,\tau) 
    & = 
        - 
        \dfrac{1}{(2\pi)^3 r}
        \dfrac{\partial}{\partial r}
        \dfrac{1}{\sqrt{\tau^2 + r^2 - t^2}}
        \ln
        \left(
            \dfrac{   |\tau| + \sqrt{\tau^2 + r^2 - t^2}    }
                  { - |\tau| + \sqrt{\tau^2 + r^2 - t^2}    }
        \right)
\end{align*}

Similarly for $g_{2}(x,\tau)$:
\begin{align*}  
    g_{2}(x,\tau)
    & = 
        \dfrac{1}
              {(2\pi)^3 r      }
        \dfrac{\partial}{\partial r}
        \dfrac{1}{2}
        \times
        2
        \dfrac{1}{\sqrt{\rho^2 + \tau^2}}
        \ln 
        \left(
            \dfrac{  |\tau| + \sqrt{\tau^2 + \rho^2}    }
                  {- |\tau| + \sqrt{\tau^2 + \rho^2}    }
        \right)
    \\
    & = 
        \dfrac{1}
              {(2\pi)^3 r      }
        \dfrac{\partial}{\partial r}
        \dfrac{1}{\sqrt{\tau^2 + r^2 - t^2}}
        \ln 
        \left(
            \dfrac{  |\tau| + \sqrt{\tau^2 + r^2 - t^2}    }
                  {- |\tau| + \sqrt{\tau^2 + r^2 - t^2}    }
        \right)
\end{align*}

Clearly, $g_{1}(x,\tau) = - g_{2}(x,\tau)$, and thus, for the case of $\rho^2 = r^2 - t^2 > 0$, 
we find $g(x,\tau) = 0$.

Returning back to \eqref{eq:starting_point_g1_and_g2} with $0 \leq  r < |t|$, we write $r = \rho \sinh(\beta)$
and $t = \epsilon(t) \rho \cosh(\beta)$ where $\rho^2 = t^2 - r^2$, and thus:
\begin{align*}
    g_{1}(x,\tau)
    & =
        -
        \dfrac{1}{2(2\pi)^3 r}
        \dfrac{\partial}{\partial r}
        \int_{0}^{\infty}
            \ud l
        \int_{-\infty}^{+\infty}
            \ud \alpha
            \,\,
    \\
    & \qquad \qquad 
            \times
            \Big(
                \cos \left( l \rho \sinh(\beta + \epsilon(t) \alpha) \right)
                +
                \cos \left( l \rho \sinh(\beta - \epsilon(t) \alpha) \right)
            \Big)
            e^{- l |\tau|}
    \\
    g_{2}(x,\tau)
    & = 
        \dfrac{1}
              {2(2\pi)^3 r     }
        \dfrac{\partial}{\partial r}
        \int_{0}^{\infty}
            \ud l
        \int_{-\infty}^{+\infty}
            \ud \alpha
            \,\,
    \\
    & \qquad \qquad 
            \times
            \Big(
                \cos \left( l \epsilon(t) \rho \cosh(\alpha + \epsilon(t) \beta)  \right)
                +
                \cos \left( l \epsilon(-t) \rho \cosh (\alpha + \epsilon(-t) \beta) \right)
            \Big)
            \sin(l |\tau|)
\end{align*}
Realigning the integration bounds, one finds:
\begin{align*}
    g_{1}(x,\tau)
    & =
        -
        \dfrac{1}{(2\pi)^3 r}
        \dfrac{\partial}{\partial r}
        \int_{0}^{\infty}
            \ud l
        \int_{-\infty}^{+\infty}
            \ud \alpha
            \,\,
            \times
            \cos \left( l \rho \sinh(\alpha) \right)
            e^{- l |\tau|}
    \\
    g_{2}(x,\tau)
    & = 
        \dfrac{1}
              {(2\pi)^3 r     }
        \dfrac{\partial}{\partial r}
        \int_{0}^{\infty}
            \ud l
        \int_{-\infty}^{+\infty}
            \ud \alpha
            \,\,
            \times 
            \cos \left( l \rho \cosh(\alpha )  \right)
            \sin(l |\tau|)
\end{align*}

Once again, after integrating first over $l$ we find:
\begin{align*}
    g_{1}(x,\tau)
    & =
        -
        \dfrac{1}{2}
        \dfrac{1}{(2\pi)^3 r}
        \dfrac{\partial}{\partial r}
        \int_{-\infty}^{+\infty}
            \ud \alpha
            \,\,
            \times
            \left[
                \dfrac{         1                   }
                      {|\tau| - i \rho \sinh(\alpha)}
                +
                \dfrac{         1                   }
                      {|\tau| + i \rho \sinh(\alpha)}
            \right]
    \\
    & = 
        -
        \dfrac{1}{2i}
        \dfrac{1}{(2\pi)^3 r}
        \dfrac{\partial}{\partial r}
        \int_{-\infty}^{+\infty}
            \ud \alpha
            \,\,
            \times
            \left[
                \dfrac{         1                   }
                      {\rho \sinh(\alpha) -i |\tau| }
                -
                \dfrac{         1                   }
                      {\rho \sinh(\alpha) + i|\tau| }
            \right]
    \\
    g_{2}(x,\tau)
    & = 
        \dfrac{1}{2}
        \dfrac{1}
              {(2\pi)^3 r     }
        \dfrac{\partial}{\partial r}
        \int_{-\infty}^{+\infty}
            \ud \alpha
            \,\,
            \times 
            \left[
                \dfrac{1}{\rho \cosh(\alpha) + |\tau|}
                -
                \dfrac{1}{\rho \cosh(\alpha) - |\tau|}
            \right]
\end{align*}

We can now do the $\alpha$ integration:
\begin{align*}
    g_{1}(x,\tau)
    & =
        -
        \dfrac{1}{2i}
        \dfrac{1}{(2\pi)^3 r}
        \dfrac{\partial}{\partial r}
        \times
        2 
        \times
        \dfrac{         (-1)                }
              {                     
                  \sqrt{(i|\tau|)^2 + \rho^2}
              }
        \ln
        \Big(
            \dfrac
            {
                i |\tau| + \sqrt{(i|\tau|)^2 + \rho^2}
            }
            {
               -i |\tau| + \sqrt{(i|\tau|)^2 + \rho^2}
            }
        \Big)
    \\
    g_{2}(x,\tau)
    & = 
        \dfrac{1}{2}
        \dfrac{1}
              {(2\pi)^3 r     }
        \dfrac{\partial}{\partial r}
        2 
        \times
        \dfrac{         1                  }
              {                     
                  \sqrt{(|\tau|)^2 - \rho^2}
              }
        \ln
        \Big(
            \dfrac
            {
                |\tau| + \sqrt{(|\tau|)^2 - \rho^2}
            }
            {
               -|\tau| + \sqrt{(|\tau|)^2 - \rho^2}
            }
        \Big)
\end{align*}

Now, we have $2$ cases:
\begin{itemize}
 \item $\rho^2 -  \tau^2 = t^2 - r^2 - \tau^2> 0$, i.e., the 5D timelike region.
 \item $t^2 - r^2 - \tau^2 < 0$, which is part of the 5D spacelike region, since we still have $t^2 - r^2 > 0$.
\end{itemize}

Let us first consider the \emph{second case}, namely, $t^2 - r^2 - \tau^2 < 0$. We then find:

\begin{align*}
    i |\tau| \pm \sqrt{(i|\tau|)^2 + \rho^2}
    & = 
        i |\tau| + \sqrt{- \tau^2 + \rho^2}
    =
        i |\tau| + \sqrt{(-1)(\tau^2 - \rho^2)}
    =
    \\
    & = 
        i ( |\tau| + \sqrt{\tau^2 - \rho^2})
\end{align*}
In which case, once again, one finds 
\begin{align*}
    g_{1}(x,\tau)
    & =
        -
        \dfrac{1}{(2\pi)^3 r}
        \dfrac{\partial}{\partial r}
        \dfrac{          1                  }
              {                     
                  \sqrt{\tau^2 + r^2 - t^2 }
              }
        \ln
        \Big(
            \dfrac
            {
                |\tau| + \sqrt{\tau^2 + r^2 - t^2 }
            }
            {
               -|\tau| + \sqrt{\tau^2 + r^2 - t^2 }
            }
        \Big)
    \\
    g_{2}(x,\tau)
    & = 
        \dfrac{1}
              {(2\pi)^3 r     }
        \dfrac{\partial}{\partial r}
        \dfrac{         1                  }
              {                     
                  \sqrt{\tau^2 + r^2 - t^2 }
              }
        \ln
        \Big(
            \dfrac
            {
                |\tau| + \sqrt{\tau^2 + r^2 - t^2 }
            }
            {
               -|\tau| + \sqrt{\tau^2 + r^2 - t^2 }
            }
        \Big)
\end{align*}
in which case, once again, $g_{1}(x,\tau) = - g_{2}(x,\tau)$. 

On the other hand, in the 5D timelike region, we have $t^2 - r^2 - \tau^2 > 0$, in which case,
the numerator and denominator of the $\ln$ argument in both $g_{1}$ and $g_{2}$ are \emph{complex conjugates}.

Let us use a shortened notation, in which $a = |\tau|$ and $b = \sqrt{t^2 - r^2 - \tau^2}$.
For $g_{1}$, we find the argument of the $\ln$ to be:
\begin{align*}
    \dfrac{i a + b}{-ia + b}
    & = 
        e^{i \tan^{-1}(a/b) - i \tan^{-1}(-a/b) } 
    =
        e^{ 2i \tan^{-1}(a/b) }
\end{align*}
Similarly, for $g_{2}$:
\begin{align*}
    \dfrac{a + ib}{-a + ib}
    & = 
        \dfrac{b - ia}{b + ia}
    =
        e^{i \tan^{-1}(-a/b) - i \tan^{-1}(a/b) } 
    =
        e^{ -2i \tan^{-1}(a/b) }
    =
        e^{2i\pi - 2i \tan^{-1}(a/b) }
\end{align*}
Thus, in the shortened notation, and recalling that $\sqrt{\tau^2 + r^2 - t^2} = i \sqrt{t^2 - r^2 - \tau^2} = ib$ we find:
\begin{align*}
    g_{1} 
    & = 
        - 
        \dfrac{1}{(2\pi)^3 r}
        \dfrac{\partial}{\partial r}
        \dfrac{(-1)}{ib} 
        2i \tan^{-1} \left( \dfrac{a}{b} \right)
    \\
    g_{2}
    & =
        \dfrac{1}{(2\pi)^3 r}
        \dfrac{\partial}{\partial r}
        \dfrac{1}{ib} 
        \left[
            2 i \pi
            -
            2i \tan^{-1} \left( \dfrac{a}{b} \right)
        \right]
\end{align*}
Clearly, when summing $g = g_{1} + g_{2}$, the $\tan^{-1}(\ldots)$ terms cancel, and we find:
\begin{align*}
    g(x,\tau)
    & = 
        \dfrac{1}{2\pi^2 }
        \dfrac{\partial}{\partial r^2}
        \dfrac{\theta(t^2 - r^2 - \tau^2)}{\sqrt{t^2 - r^2 - \tau^2}}
\end{align*}
which is, once again, the conventional solution.

Thus, even for this method of integrating $k_{5}$ first, we have obtained \eqref{eq:green_function_expected_solution}
(a different result was obtained in \cite{OronHorwitz2000} due to an \emph{error in sign}).

\section{Conclusions and discussion}
We have shown that the 5D Green Function is reproduced with the same methods 
used in \cite{OronHorwitz2000} and \cite{LandHor1991}, 
showing that the different methods used lead to equivalent results.
In this, we believe that the apparent form of the Green function discrepancy has been removed,
and one can utilize the \emph{$\tau$-retarded} conventional Green function for computing the fields.
In \cite{aharonovich_2009}, we have used the method of Nozaki \cite{Nozaki_1964},
who derived generalized fundamental solutions for the $O(p,q)$  wave equation,
to obtain an explicit $\tau$-retarded Green function.
The form of the Green function has direct implications on the form of the fields produced by charges,
and in particular, on the problem of radiation reaction.

We have chosen the $\tau$-retarded Green-function in this work,
and for future applications, in accordance with what appears to be the dynamical
structure of the 5D fields. In contrast to the solutions of the standard Maxwell theory,
where the $t$ variable is subject to the action of the Lorentz transformation 
as a \emph{fundamental symmetry}, the $\tau$ variable does not necessarily 
participate in a higher symmetry.
Although the equations for the fields have $O(4,1)$ symmetry, in this case, the manifestly covariant dynamics
of the sources evolve by the $O(3,1)$ invariant $\tau$ serving as a \emph{universal evolution parameter}.

Non-relativistic mechanics is conventionally considered to evolve according to a Galilean invariant parameter $t$
(which may be identified with the parameter $\tau$ of the relativistic theory), corresponding to Newton's
basic postulate of a \emph{universal time}.
The relativistic dynamics of Stueckelberg \cite{Stueckelberg1941,Stueckelberg1942} considers 
the events $x^{\mu}(\tau)$ to be the basic dynamical elements of the relativistic theory of matter.
Jackson's construction \cite{Jackson1995} for a covariant current is an example for this interpretation
(see also \cite{LandShnerbHorwitz1995,SaadHorArsh1989}).
The quantization of the 5D fields that arise as the $U(1)$ gauge fields from the Stueckelberg-Schr\"{o}dinger equation \cite{Horwitz1993} 
(see eq. \eqref{eq:stueckelberg_hamilton_equation})
indicate that these fields evolve in $\tau$, unlike the Maxwell fields which have no
distinguished evolution parameter.
One may therefore impose a causal structure on the fields governed by this distinguished parameter $\tau$,
through the use of the $\tau$-retarded propagator.

Application of the explicit $\tau$-retarded solution to the radiation reaction problem 
will be reported in a succeeding publication.

\newpage
\bibliographystyle{amsplain}
\bibliography{bibliography-all}

\end{document}